\documentstyle[11pt,aaspp4]{article}

\tighten   

\begin{document}

\title {CHEMICAL ABUNDANCE CALIBRATIONS FOR THE NARROW-LINE-REGION OF ACTIVE  GALAXIES}

\author{Thaisa Storchi-Bergmann}
\affil{Departamento de Astronomia, IF-UFRGS,
	CP\,15051, CEP\,91501-970,
	Porto Alegre, RS, Brasil}

\author{Henrique R. Schmitt}
\affil{Departamento de Astronomia, IF-UFRGS,
	CP\,15051, CEP\,91501-970,
	Porto Alegre, RS, Brasil}

\author{Daniela Calzetti}
\affil{Space Telescope Science Institute,
	3700 San Martin
	Dr., MD 21218}

\author{Anne L. Kinney}
\affil{Space Telescope Science Institute,
	3700 San Martin
	Dr., MD 21218}

\begin{abstract}

We investigate two chemical abundance calibrations for the narrow-line-region
(NLR) of active galaxies in terms of three easily observable optical
emission-line ratios, namely,
[OIII]$\lambda\lambda$4959,5007/H$\beta$, 
[NII]$\lambda\lambda$6548,84/H$\alpha$ and
[OII]$\lambda3727/$[OIII]$\lambda\lambda$4959,5007.
The calibrations are obtained from a grid of models
on the assumption that the main process responsible for
the production of these lines is photoionization by a ``typical''
active galactic nucleus continuum. The chemical elements vary their abundance 
together with oxygen, except nitrogen, which is 
assumed to be a product of secondary nucleosynthesis.
The calibrations are calculated for the 
range $8.4 \le 12+log(O/H) \le 9.4$, and tested
using NLR data for a sample of Seyfert's and LINER's having 
HII regions in the vicinity of the nucleus. The gaseous abundances
of these HII regions have been determined in previous works,
and the NLR abundances are obtained on the assumption that
they can be extrapolated from those of the HII regions.
The calibrations work very well for the Seyfert's,
giving abundance values which agree with those obtained from
the HII regions, and can thus be used for quick estimates of the chemical
abundances of the NLR's. 
For the LINER's, the calibrations give lower values
than those derived from the corresponding HII regions, suggesting that the
assumptions of the models do not apply for them, and that there are
different physical processes at work in the NLR of the LINER's.


\end{abstract}

\section{Introduction}

Since the works of Ferland \& Netzer (1983) and Halpern \&
Steiner (1983), on the modelling of the the narrow-line-region (NLR)
of active galactic nuclei (AGN), which were
followed by a large number of subsequent works (e. g. Stasi\'nska
1984; Ferland \& Osterbrock 1986; Binette, Robinson \& Curvosier 1988),
it has been shown that the relative fluxes of the stronger emission-lines 
can be successfully reproduced assuming that the main excitation 
mechanism of the NLR is photoionization.
Alternative photoionization models include varying proportions
of matter-bounded and ionization-bounded clouds in the NLR
(Viegas-Aldrovandi 1988; Binette, Wilson \& Storchi-Bergmann 1996),
which are better suited to reproduce both the high and low excitation
emission-line ratios. 
Viegas-Aldrovandi \&  Contini (1989) have also investigated the
effect of shocks, concluding that clouds with shock velocities
lower than 300~km~s$^{-1}$ are radiation-dominated if the ionization
parameter is larger than U=10$^{-4}$. Viegas-Aldrovandi
\& Gruenwald (1990) have also shown that, besides shocks,
relativistic particles have to be taken into account
to reproduce the spectra of LINER's. 
More recent studies of the effect of shocks
in the NLR or ENLR (extended narrow-line region -- expression sometimes
used to indicate that the NLR is obivously extended, usually to kpc
scales) include the
works of Viegas and Gouveia dal Pino (1992),
Sutherland et al. (1993) and Dopita \& Sutherland (1995).

The conical shape observed for the ENLR in a number of
Seyfert galaxies seems to favor photoionization by a nuclear
source as the ionizing mechanism (Wilson, 1995 and references
therein). This is confirmed by long-slit spectroscopy along
the extended gas for a few Seyferts, like NGC~3281 (Storchi-Bergmann,
Wilson \& Baldwin 1992), Mrk~573 (Tsvetanov \& Walsh 1992)
and NGC~5643 (Schmitt, Storchi-Bergmann \& Baldwin 1994).
The variation of the emission-line ratios with distance from the
nucleus in these three galaxies can be reproduced by a sequence of
models with varying ionization parameter, even when there is evidence
of the presence of shocks with velocities $\sim$150~km~s$^{-1}$
in the gas (e.g. NGC~3281).

Most works devoted to reproducing the  emission-line ratios
of the NLR or ENLR via  photoionization models, 
adopt a solar composition for the gas. 
Collecting NLR emission-line ratios from a sample
of 180 LINER's and Seyfert 2's, Storchi-Bergmann
\& Pastoriza (1990), have shown that the data
is better reproduced if a range of nitrogen
abundances, from solar to 4 times solar, is adopted.

In order to verify this result, Schmitt et al. (1994), Storchi-Bergmann,
Wilson \& Baldwin (1996a), Storchi-Bergmann et al. (1996b) 
have obtained long-slit
spectroscopy of a sample of AGN with HII regions
in the vicinity of the nucleus, such that the abundances
could be determined from the HII regions spectra instead
of using models for the NLR (on the assumption that the nuclear,
or NLR abundance, is the same as that of the HII regions, 
due to their proximity to the nucleus).
The derived nuclear oxygen abundances
range from solar to two-three times solar, and
the nitrogen  shows a secondary behavior,
such that its abundance can reach up to 4-5 times solar,
in accordance with the previous results from the modelling of
the NLR. 

Nevertheless, the number of active galaxies for which such determinations
are available is small, 
and there are also several active galaxies which lack
HII regions close to the nuclei, such that the above method cannot be used. 
In this work, we use the HII region abundances derived
in previous works, to infer that of the NLR and
constrain, via a grid of photoionization 
models, the average physical parameters of the  NLR.
We then  investigate the relation between the emission-line-ratios and
chemical abundances. Under a few assumptions, discussed
below, we obtain two calibrations
which allow the derivation of the chemical abundance
of the NLR gas in terms of two easily observed optical emission-line ratios.

\section{Calculations}

The grid of models was constructed using the code CLOUDY (Ferland 1996),
for three different ionizing continua: two power-laws 
F$_\nu\propto\nu^{-\alpha}$, for $\alpha=$1.0 and 1.5
(turn-on and turn-off energies of $0.1 eV$ and $100 keV$, respectively) and
the option ``Table AGN'' in Cloudy, which 
uses a typical observed AGN continuum (a combination
of power-laws; see Mathews and Ferland 1987). 
The gas densities were varied between $10^2\le N \le 10^4$ cm$^{-3}$,
as previous works (e.g. Schmitt et al. 1994;
Storchi-Bergmann et al. 1992; Tsvetanov \& Walsh 1992) indicate that
most observed NLR gaseous densities are in this range.
The ionization parameter was varied in the interval
$-4.0 \le log(U)\le -2.0$, which allows the reproduction of
the observed range in emission line ratios for typical NLR's.

As pointed out in the introduction, the oxygen abundances obtained
in previous works for the HII regions close to the nuclei of Seyfert 
galaxies range from about solar to a few times solar, and we have assumed
that the NLR's have similar abundance values. The oxygen
abundance was thus varied in the range  $8.4 \le 12+log(O/H)\le 9.4$.
The adopted solar abundance values are
$12+log(O/H)=8.91$ and $12+log(N/H)=7.98$, from Aller(1987) and 
Grevesse(1984).
All elements, except nitrogen, are assumed to have abundances
equal to that of oxygen, relative to the solar value.

Nitrogen was assumed to behave as a secondary element.
The main observational constraint on the origin of nitrogen
is the behavior of the 
abundance ratio N/O as a function of the overall metallicity,
as measured by the O/H ratio.  In the ``simple
chemical evolution model" 
(Edmunds 1990), N/O will be a constant
for a primary origin of nitrogen, whilst N/O will be proportional 
to O/H for a secondary origin. Vila-Costas \& Edmunds (1993) 
have used HII region data for a number of spiral and irregular
galaxies to reach the conclusion that the N/O and O/H ratios
can be consistently explained by models in which nitrogen
has mostly a secondary origin at high abundances, namely 
for $12+log(O/H) > 8.5$. We have reached a similar conclusion
analysing the data of a number of nuclear starbursts
(Storchi-Bergmann, Calzetti \& Kinney 1994), and
of HII regions in the vicinity of active nuclei (Storchi-Bergmann
et al. 1996a,b). As we intend to obtain a calibration
for the NLR, which is located in the nuclear region of 
luminous and metal rich galaxies we will thus assume
that nitrogen has a secondary behavior also in the NLR,
with abundance given by the relation:

\begin{equation}
log(N/O)=0.96\times[12+log(O/H)]-9.29
\end{equation}

\noindent
which was obtained by Storchi-Bergmann et al. (1994) 
for nuclear starbursts.

Another aspect which has also become evident in recent studies 
(e.g. Schmitt et al. 1994;
Storchi-Bergmann et al. 1992; Mulchaey, Wilson \& Tsvetanov 1996)
is that the nuclear region of Seyfert~2 galaxies is obscured, denoting the 
presence of dust in the NLR. 
The presence of dust was considered using the command ``grains'' in
CLOUDY, and depleting the heavy elements abundances as follows: for
a solar abundance nebula, the abundance of the depleted
gas, relative to the solar value is:
He$=$1.0, C$=$0.39, N$=$0.81, O$=$0.62, Ne$=$1.0, 
Mg$=$0.26, Al$=$0.01, Si$=$0.045, S$=$0.59, Ar$=$0.44, Ca$=$0.00009, 
Fe$=$0.0089, Na$=$0.024 and Ni$=$0.01.
These values correspond to the observed abundance of the
interstellar medium (Cowie \& Songaila 1986).
For chemical compositions different than solar, the depletions were
scaled accordingly.

\section{Discussion}

Under the above assumptions, 
the parameters which determine the emission-line ratios 
of the NLR are: the spectrum of the ionizing continuum,
the ionization parameter, the chemical abundance and the gaseous
density. 

The most easily observed optical emission-lines
from the NLR of Seyferts
are [OII]$\lambda$3727 (hereafter [OII]), H$\beta$, [OIII]$\lambda\lambda$4959,5007 (hereafter [OIII]), H$\alpha$,
[NII]$\lambda\lambda$6548,84 (hereafter [NII])
and [SII]$\lambda\lambda$6717,31
(hereafter [SII]).
>From these, a number of line ratios can be obtained, for example:
[OII]/[OIII], which is independent of the chemical abundance and is
a good indicator of the ionization parameter;
[OIII]/H$\beta$, which depends on all the parameters,
but also frequently used as an indicator of the ionization parameter;
[NII]/H$\alpha$, mostly sensitive to the gaseous abundance; [SII]$\lambda$6717$/\lambda$6731, which is
sensitive to the density.

If we adopt the simplifying assumption that the density 
is indicated by the [SII] ratio, and that the 
ionizing spectrum is also known, then we can choose two other
emission line ratios as indicators of the ionization parameter and chemical
abundance. From the theoretical point of view, the ratio
[OII]/[OIII] is the best choice for the derivation of the ionization
parameter, as it does not depend on the chemical abundance of the gas.
But, from the observational point of view, it has the disadvantage of
being sensitive to the reddening. In order to avoid this problem,
the ratio [OIII]/H$\beta$ can be used instead, but it will be also
sensitive to the abundance, and should be used together with the ratio
mostly sensitive to the abundance, which is
[NII]/H$\alpha$. These last two ratios have the advantage of not being
sensitive to reddening.

In order to derive a chemical abundance calibration for the NLR,
we have explored two diagrams: $[NII]/H\alpha \times [OIII]/H\beta$
and $log([OII]/[OIII]) \times log([NII]/H\alpha)$.
In Fig. 1 we present a grid of models in the plane 
$[NII]/H\alpha \times [OIII]/H\beta$, for a range of oxygen abundances
$8.4\le 12+log(O/H) \le 9.2$. [Hereafter, we will refer to
12+log(O/H) as (O/H).] In these models, the gas density is
300 cm$^{-3}$ (typical for the NLR) and the ionizing continuum is
the segmented power-law of Mathews \& Ferland (1987) (typical
of AGN). In Fig. 2, the same grid of models is plotted in
the plane $log([OII]/[OIII]) \times log([NII]/H\alpha)$, for a
range of oxygen abundances $8.4\le 12+log(O/H) \le 9.4$.

It can be concluded from Figures 1 and 2 that 
a pair of ratios [OIII]/H$\beta$, [NII]/H$\alpha$ 
or $log$([OII]/[OIII]), $log$([NII]/H$\alpha$) can 
determine an abundance value. Notice the clean sequencing
of models in terms of the abundance values in the figures,
specially in the latter diagram (Fig. 2). This diagram is
particularly recommended for high abundances [(O/H)$>$9.2] 
because of some degeneracy of the models 
in the first diagram (Fig. 1). 

We have then fitted two-dimensional second-order polinomials 
to each of the diagrams in order to derive a
calibration for the oxygen abundance in terms
of the above emission-line ratios.
For the first diagram (Fig. 1) we obtain, for $x=[NII]/H\alpha$
and $y=[OIII]/H\beta$, interpolating in the interval 8.4 $\le (O/H) \le$ 9.2:

\begin{equation}
(O/H)=8.34+0.212x-0.012x^2-0.002y+0.007xy
-0.002x^2y+6.52\times 10^{-4}y^2+2.27\times 10^{-4}xy^2+8.87\times 10^{-5}x^2y^2
\end{equation}

For the second diagram, the best fit obtained for 
$u=log([OII]/[OIII])$ and $v=log([NII]/H\alpha$ in the interval 
8.4 $\le (O/H) \le$ 9.4 is:

\begin{equation}
(O/H)=8.643-0.275u+0.164u^2+0.655v-0.154uv
-0.021u^2v+0.288v^2+0.162uv^2+0.0353u^2v^2
\end{equation}

\noindent
For both expressions, the fitted values are within 
0.05 dex of the model values, with
$\chi^2$ of $1.04\times 10^{-3}$ and $6.9\times 10^{-4}$, respectively.

The dependence of the calibrations on the density can be quantified as
far as the density stays approximately within the range
$100\le N\le 10000$ cm$^{-3}$. 
For larger densities, [NII]/H$\alpha$ is not a good abundance indicator 
anymore, because it is suppressed as its critical density is
8.6$\times$10$^4$ cm$^{-3}$, and the calibrations are not valid.
As the density is increased,
both [OIII]/H$\beta$ and [NII]/H$\alpha$ ratios increase
sistematically, and the dependece of
the two calibrations on the density 
is approximately linear in the logarithm of the density.
This dependence can be incorporated 
into our calibrations, as:

\begin{equation}
(O/H)_{final}=(O/H)-0.1 log(N/300)
\end{equation}

\noindent
where $N$ is the gaseous density in cm$^{-3}$, and the equation is valid
for $100\le N\le 10000$ cm$^{-3}$. 

Calibrations using power-law ionizing 
continua  $F_\nu\propto\nu^{-\alpha}$, for $\alpha=-1$ and -1.5. were also obtained. As compared with the above calibrations, for which we used the
typical AGN continuum of Mathews and Ferland (1987),
it can be concluded that they give sistematically
larger values for the calculated abundances.
The second calibration (in terms of $log([NII]/H\alpha$ and 
$log([OII]/[OIII])$) is less dependent on the
ionizing continuum, giving abundance values from 0.1 to 0.3 $dex$
larger with the power-law than with the AGN continuum.
For the first calibration, the difference is increased, reaching
values larger than 0.5$dex$. We point out that, 
although the shape of the ionizing
continuum in AGN is not very well stablished, UV and X-ray observations
of at least the brightest AGN show that it is closer to a
composition of power-laws than to a unique power-law
(e. g. Sanders et al. 1989; Alloin et al. 1995), favoring
the ``AGN continuum'' used above over the single power-laws.

\section{Comparison with Observations}

In order to verify the validity of the above calibrations,
we have collected from the literature the emission line ratios
of the NLR of a few Seyfert's and LINER's, for which there are determinations
of the oxygen and nitrogen abundances for HII regions close
to the NLR. In this way, the NLR abundances could be 
extrapolated from those of the HII regions. 

The collected data are shown in Table 1, along with the the
Hubble type and absolute magnitude for each galaxy. 
The emission-line fluxes are listed relative to H$\beta$ and 
have been corrected for reddening, and their source is listed in the
last column of the table. Also listed are the electronic densities
calculated from the [SII]$\lambda\lambda$6716/6731 line ratios 
as in Osterbrock (1989; Fig. 5.3). In applying the correction 
to the (O/H) due to the density dependence (equation 4), we
will assume that the gas density is approximately the same as
the electronic density (N$\approx$N$_e$).

Typical radial abundance gradients
for normal galaxies range between 0.05 and 0.1 dex kpc$^{-1}$ -- see, for 
example, Vila-Costas \& Edmunds (1992) 
and Kennicutt et al (1993). Schmitt et al. (1994), 
Storchi-Bergmann et al. (1996a,b)
have shown that, for the Seyfert~2 NGC~5643 and the LINER's 
NGC~1097, NGC~1672 and NGC~1598, the observed gradients are similar
to those typical of normal galaxies with the same morphological type.
Here we thus adopt the latter hypothesis in the extrapolation of
the NLR abundances, for the cases in which a gradient is not available.

In Table 2 we present the resulting (O/H) values of the
NLR's obtained with each of the proposed calibrations
-- equations 2, 3 and 4 above, together with
the derived values from the extrapolation using the 
abundances of the HII regions. We also show, in the
last two columns of the table, the adopted gradient in (O/H)
and the references from which the HII regions' abundances
and gradients were obtained. 
Errors in the calibrations have been calculated from the
errors in the emission-line ratios, while errors in the
extrapolated values have been estimated from the errors
in the HII regions abundances and gradients.

The O/H values obtained from the two calibrations have
been plotted against the ones extrapolated from the HII regions in 
Figure 3. It can be seen that, within the errors,
there is a reasonable agreement between the extrapolated 
and calculated values for the
Seyfert 2's, with maximum differences of $\sim$0.2 dex between the calculated and extrapolated values, indicating that the calibration works
very well for the Seyfert 2's.

The (O/H) values obtained using the calibrations can also be
compared with previous determinations via detailed modellling
of the NLR, which we have for NGC\,5643 (Schmitt et al. 1994)
and another Seyfert 2 galaxy, NGC\,3281 (Storchi-Bergmann et al. 1992).
In both cases, the models indicated solar abundance for the
gas (O/H=8.91), while the calibrations give (O/H)=9.00 (equations
2 and 4) and (O/H)=8.94 (equations 3 and 4) for NGC\,5643 and
(O/H)=8.81 and 9.03, respectively, for NGC\,3281, which confirm
the validity of the calibrations.

On the other hand, for all the 4 LINER's the 
calibrations give values which are
systematically lower than those of the Seyfert's.
But we note that Storchi-Bergmann et al. (1996a,b) 
have concluded that the LINER nucleus of NGC~1672 
presents emission line ratios which can be reproduced with photoionization
by hot stars, and the absorption spectra from both this nucleus and that of 
NGC~1598 present signatures of young stars.
>From the discussion above, the calibrations would not be valid
for these two cases, identified by open triangles in Fig. 3.
But they should be valid for the other two LINER nuclei.
As there is no systematic difference between the extrapolated 
nuclear abundances
for the NLR's of Seyfert's and LINER's, the different behavior
of the LINER's in the diagrams suggests a difference 
between the physical conditions and/or ionizing source
in these objects as compared with the Seyfert's.
One possibility is that the energy distribution
responsible for the photoionization of the gas in
LINER's is an absorbed continuum.
Matter-bounded clouds from the BLR (Ferland et~al. 1996) or
from the NLR (Binette et~al. 1996) would be responsible for the
absorption. Nevertheless, our results are based on observations of
only two objects. More LINER's should be subject of this kind of study
before a firm conclusion can be reached.

We also show in Figure 4, the results of the two calibrations
plotted against each other. Although the correlation is not perfect, 
(as the calibrations are fits to the sequences of models), 
the Spearman correlation coefficient between the two calibrations
is $r_s=0.80$, indicating a good correlation.
 It can also be noticed that there is a sistematic
shift between the two calibrations, such that the second gives (O/H)
values on average 0.11$dex$ larger than the first calibration.
As judged from the comparison with the observations, the (O/H) value to
be adopted when both calibrations can be used is the average from 
the two. 

As pointed out in the previous section, we have also calculated 
calibrations for power-law ionizing continua instead of the
``AGN continuum'' used to obtain the calibrations of equations 2 and 3.
Nevertheless, the comparison of the calculated (O/H) with the 
observed ones (from the HII regions) give a much poorer agreement.
The calibration in terms of log([NII]/H$\alpha$) and log([OII]/[OIII])
(second calibration) gives a somewhat better result than  
the one in terms of [NII]/H$\alpha$ and [OIII]/H$\beta$ (first calibration):
although the calculated (O/H) are sistematically
overestimated when compared with the
observed ones, the difference stays within the range 0.1 to 0.3$dex$.
But the first calibration gives (O/H) values larger than the observed by
0.5$dex$ in some cases. These results suggest that the 
``AGN continuum'' is a better representation of the ionizing continuum
than the single power-laws for the sample galaxies.

It is important to point out that there may be at least 
one source of confusion in observed 
emission line ratios from the NLR: the proximity of HII regions,
such that the observed emission-line ratios
could be due to a mixture of gas photoionized by an AGN continuum
and by blue stars. In principle,
such cases could be sorted out using diagnostic
diagrams (Baldwin, Phillips \& Terlevich, 1981; 
Veilleux \& Osterbrock 1987). The calibration is valid
only for those cases in which the emission-line ratios
are clearly located in the region of the diagrams corresponding to
the Seyferts. In the ``mixed '' cases, the line ratios
are intermediate between those of Seyfert's and those of 
HII regions, and the calibration is not valid.

\section{Summary and Final Remarks}

We have collected chemical abundance data of HII regions in the vicinity of the
nuclei for  a sample of 11 active galaxies and have used them to extrapolate the
chemical abundances of the corresponding NLR's. These abundances were
used to test calibrations which
allow the determination of the chemical abundance of the NLR
of Seyfert galaxies in terms of two easily observed optical emission-line ratios. Two calibrations were obtained, the first
involving a linear combination of the ratios
[NII]/H$\alpha$ and [OIII]/H$\beta$ and the second,
a linear combination of the decimal logarithms of
[NII]/H$\alpha$ and [OII]/[OIII]. Although the
first calibration involve ratios less sensitive to
the reddening, we have concluded that the second calibration
is less sensitive to the ionizing continuum spectrum.
Thus, whenever possible, both calibrations should be used
(equations 2, 3 and 4) and the (O/H) abundance 
calculated as the average between the two values.
The calibrations work well for  all the Seyfert's,
suggesting that the  hypotheses in the
modelling of the NLR are valid for most Seyfert's.

The calibration seems not to work for the LINER's,
suggesting that different ionization mechanisms occur 
in these objects. Nevertheless, more LINER's should be
studied before a firm conclusion can be reached, as
there are only 4 LINER's in the sample, and
for 2 of them there seems to be contamination of
the NLR by surrouding HII regions, and the calibrations do not
work in these cases. 

It would be important to test the calibrations in more
objects, with a larger abundance range,
but presently there are no other active galaxies for which there
are determinations of the chemical abundance of their HII regions in
the literature.
A first step to look more closely into the abundances of HII
regions in a larger number of Seyferts
has been made by Evans et al. (1996), who have recently published an
atlas with the optical positions of HII regions in 17 Seyferts.
The following step would be to make spectroscopic
observations of the individual HII regions.

In summary, we point out that determining the chemical abundance
of the NLR in Seyfert's is not an easy task. 
In previous works we have explored two methods:
modelling of individual NLR and abundance determination 
of HII regions in the vicinity of the NLR. In this work we propose
an alternative method: simple calibrations to be used when just 
a quick estimate of the chemical abundance is necessary.

\acknowledgments

This research received partial support of the Brazilian institutions
CNPq, Finep and FAPERGS. We acknowledge fruitful discussions with
L. Binette, and the suggestions of an anonymous referee, which helped
to improve the paper. This research has made use of the NASA/IPAC 
Extragalactic Database (NED)   
which is operated by the Jet Propulsion Laboratory, California Institute   
of Technology, under contract with the National Aeronautics and Space      
Administration.

%
%


\begin{table*}
\begin{center}
\begin{tabular}{lcrrrrrrc}
Galaxy &Type\tablenotemark{a} &M$_B$\tablenotemark{a,b} &[OII]  &[OIII]  
&H$\alpha$ &[NII] &N$_e$(cm$^{-3})$ &Ref.\tablenotemark{c} \\
{\bf Seyfert's} &&&&&&&&\\
IC~1816  &Sab &-20.3 &2.0$\pm$0.2  &17.9$\pm$0.5  &2.9$\pm$0.2  &6.9$\pm$0.3 &650 &1  \\  
NGC~1068 &(R)SA(rs)b  &-21.3    &1.2$\pm$0.2  &16.5$\pm$0.4  &2.6$\pm$0.2  &6.1$\pm$0.3 &8000 &2  \\ 
NGC~1386 &Sa &-19.0 &2.2$\pm$0.2  &15.9$\pm$0.4  &3.0$\pm$0.2  &5.0$\pm$0.3  &650 &1  \\ 
NGC~1566  &SAB(rs)bc &-21.2 &2.0$\pm$0.3  &15.0$\pm$0.4  &3.1$\pm$0.4  &4.1$\pm$0.4 &650 &3 \\ 
NGC~3081 &S0/a &-20.1 &1.5$\pm$0.2  &15.1$\pm$0.4  &3.0$\pm$0.2  &3.7$\pm$0.2 &200 &1  \\   
NGC~5643 &SAB(rs)c  &-20.3      &5.20$\pm$0.5  &16.4$\pm$0.4 &3.0$\pm$0.2   &4.2$\pm$0.2 &200 &4  \\ 
NGC~6814 &SAB(rs)bc &-19.5  &3.3$\pm$0.2 &15.8$\pm$0.4 &2.8$\pm$0.2 &6.0$\pm$0.3 & 650 &5 \\
{\bf LINER's} &&&&&&&\\
NGC~1097  &SB(rs)bc      &-21.0      &4.0$\pm$0.4   &5.1$\pm$0.3   &2.9$\pm$0.2  &7.2$\pm$0.4 &400 &6\\
NGC~1326  &RSBa  &-19.7 &13.2$\pm$0.6  &2.6$\pm$0.2   &2.9$\pm$0.2  &5.8$\pm$0.3 &650 &1\\
NGC~1598  &Sbc   &-20.2 &0.7$\pm$0.1   &2.3$\pm$0.2   &3.0$\pm$0.2  &4.7$\pm$0.2  &100 &1\\ 
NGC~1672  & SB(s)b     &-19.9      &1.3$\pm$0.1   &2.1$\pm$0.2   &2.9$\pm$0.2  &4.1$\pm$0.2 &250 &6\\
\tablenotetext{a}{From Storchi-Bergmann et al. 1996a, and 1996b, except for NGC~1068, NGC~1566, NGC~5643
and NGC~6814, for which we have used the NED-IPAC Extragalactic Database}
\tablenotetext{b}{H$_0=75~km~s^{-1}Mpc^{-1}$}
\tablenotetext{c}{References: (1) Storchi-Bergmann et al. 1996b; (2) Koski 1978; (3) Alloin et al. 1985; (4) Schmitt et al. 1994;
(5) Our unpublished data; (6) Storchi-Bergmann et al. 1996a} 
\end{tabular}
\end{center}

\caption{Emission-line fluxes relative to H$\beta$ (reddening-free)}

\end{table*}


\begin{table*}
\begin{center}
\begin{tabular}{lrrrcc}
Galaxy   &O/H\tablenotemark{a}  &O/H\tablenotemark{b}   &O/H\tablenotemark{c}  &$\Delta$(O/H)\tablenotemark{d}&Ref.\tablenotemark{e}  \\
{\bf Seyfert's} &&&&&\\
IC~1816    &9.35$\pm$0.09  &9.34$\pm$0.04   &9.3$\pm$0.1   &-0.06     &1 \\
NGC~1068   &9.14$\pm$0.08  &9.35$\pm$0.06   &9.1$\pm$0.1   &--  &2,3 \\
NGC~1386   &9.02$\pm$0.06  &9.15$\pm$0.04     &8.9$\pm$0.1  &-0.06   &1 \\
NGC~1566   &8.87$\pm$0.07  &9.07$\pm$0.07     &9.1$\pm$0.2  &-0.08  &4 \\ 
NGC~3081   &8.90$\pm$0.05  &9.17$\pm$0.05     &9.1$\pm$0.2  &-0.02   &1,5 \\
NGC~5643   &9.00$\pm$0.05  &8.95$\pm$0.04     &9.0$\pm$0.1  &-0.05  &6\\
NGC~6814   &9.15$\pm$0.07  &9.14$\pm$0.05    &9.3$\pm$0.2  &-0.07   &2,5 \\
{\bf LINER's} &&&&&\\
NGC~1097   &8.84$\pm$0.04     &8.97$\pm$0.04     &9.3$\pm$0.1   &-0.05 &7 \\
NGC~1326   &8.70$\pm$0.03     &8.70$\pm$0.03     &9.0$\pm$0.1   &-0.06 &1\\
NGC~1598   &8.71$\pm$0.02     &9.03$\pm$0.04     &9.2$\pm$0.1   &-0.06 &1\\
NGC~1672   &8.64$\pm$0.02     &8.83$\pm$0.03     &9.2$\pm$0.2   &-0.06 &7\\

\tablenotetext{a}{12+log(O/H) from the first calibration (eqs. 2 and 4)}
\tablenotetext{b}{12+log(O/H) from the second calibration (eqs. 3 and 4)}
\tablenotetext{c}{Extrapolating from the HII regions}
\tablenotetext{d}{Adopted gradient in dex kpc$^{-1}$}
\tablenotetext{e}{(1) Storchi-Bergmann et al. 1996b; (2) Evans \& Dopita 1987; (3) Oey \& Kennicutt 1993; (4) Hawley \& Phillips 1980;
(5) Vila-Costas \& Edmunds 1992; (6) Schmitt et al. 1994; (7) Storchi-Bergmann et al. 1996a}  
\end{tabular}
\end{center}

\caption{Oxygen abundances of the NLR}

\end{table*}

%
%


%
%

\clearpage

\begin{figure}
\caption{
Sequence of photoionization models in which the gaseous density
is 300~cm$^{-3}$, the dust-to-gas ratio is solar,
and the ionization parameter is varied in the range $-4\le log(U) \le -2.0$.
Each sequence corresponds to a different chemical abundance,
from $12+log(O/H)=8.4$ (bottom) to 9.2 (top).}
\end{figure}

\begin{figure}
\caption{Same as Fig. 1 for the plane log([OII]/[OIII]) $\times$
log([NII]/H$\alpha$), in the range $8.4\le 12+log(O/H) \le 9.4$.}
\end{figure}

\begin{figure}
\caption{The oxygen abundance values [12+log(O/H)] for the NLR
obtained from the two proposed calibrations, plotted against the
values obtained from the HII regions. Filled symbols represent 
the Seyfert 2 galaxies, and open symbols the LINER's. The
triangles represent the LINER's for which there are 
signatures of recent star formation in the nucleus. Bottom:
first calibration, involving the line ratios [NII]/H$\alpha$ and
[OIII]/H$\beta$ (Fig.1); top: second
calibration, involving $log([NII]/H\alpha)$ and
$log([OII]/[OIII])$ (Fig. 2). Both calibrations have been
corrected by a small dependence on the gas density (see text).
The loci of equal values for the two quantities are drawn as a line for comparison.}
\end{figure}

\begin{figure}
\caption{The oxygen abundance values [12+log(O/H)] obtained with the
second calibration (Fig. 2), plotted against the values 
obtained with the first calibration (Fig. 1). 
For comparison, the loci of equal values for the two quantities are drawn as a 
continuous line, and those corresponding to a sistematic shift
of the second calibration to [12+log(O/H)] values 0.11$dex$ larger, as
a dashed line. Symbols as in Fig. 3.}
\end{figure}

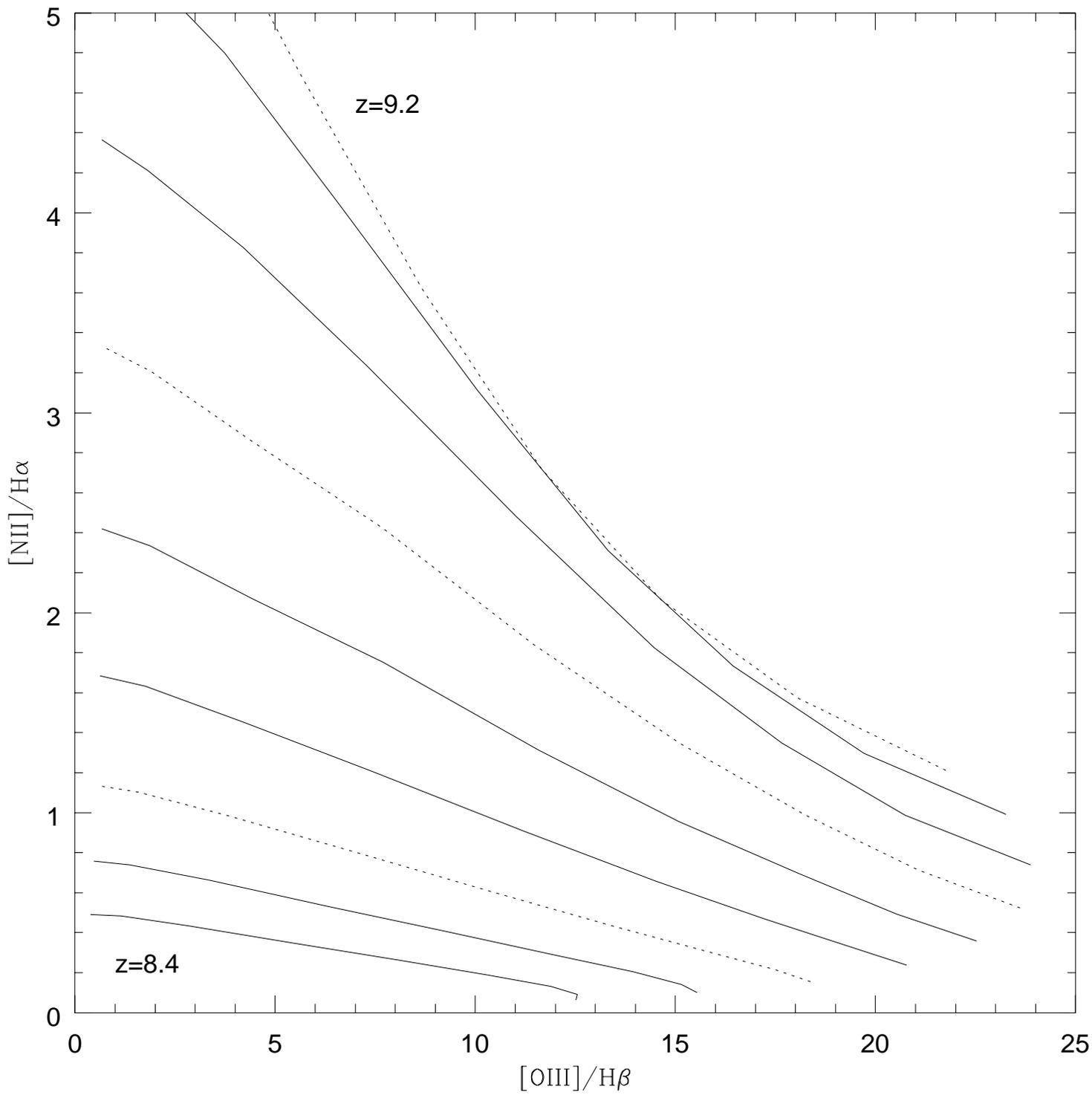
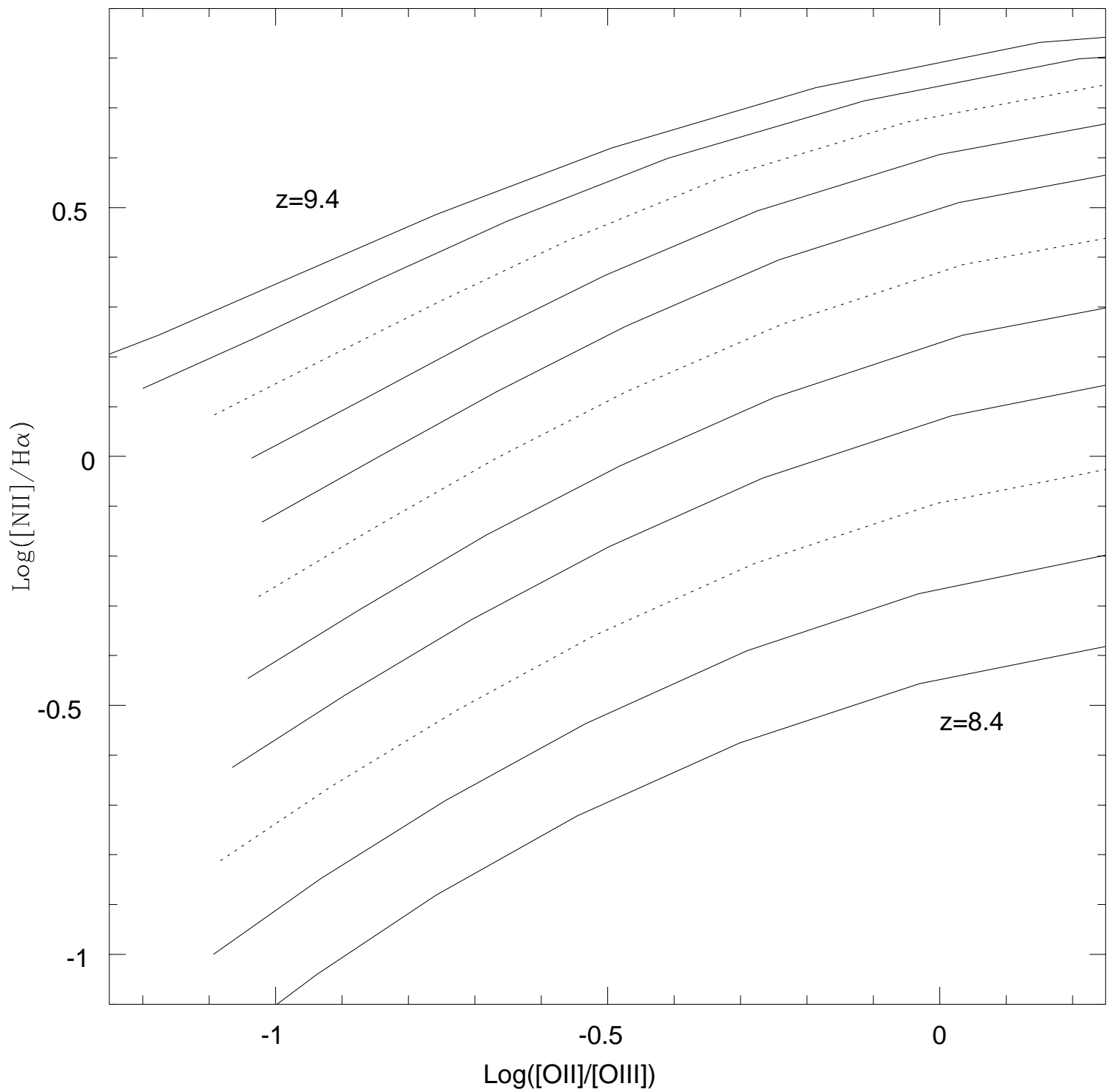
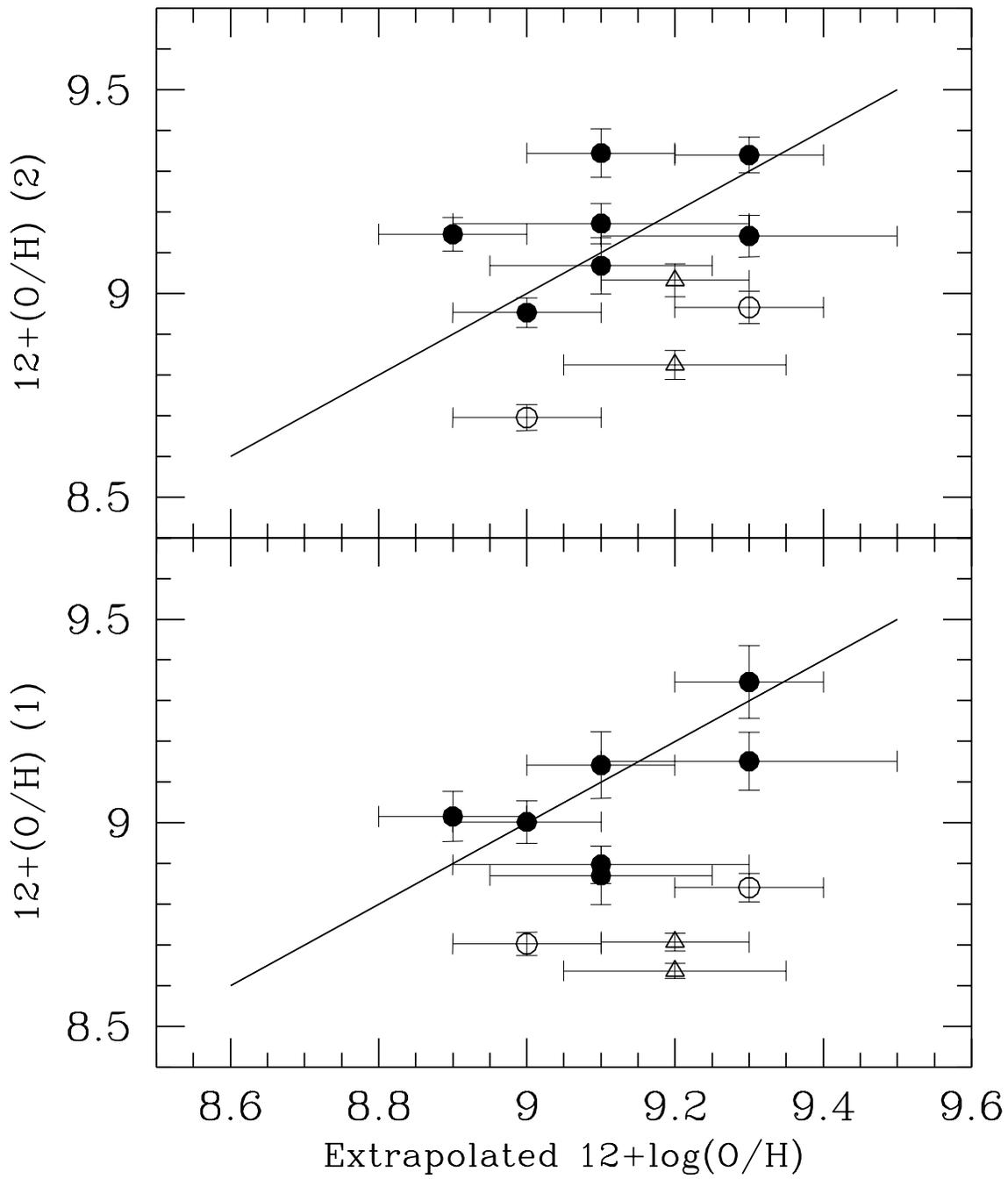
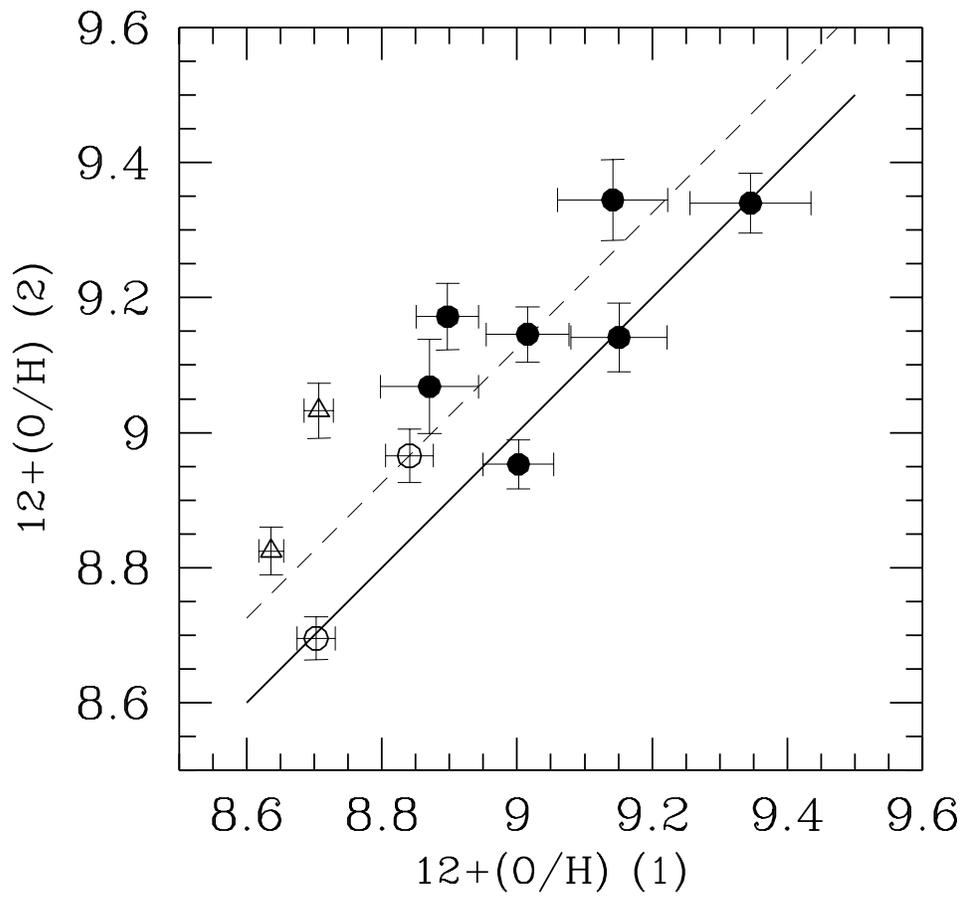
\end{document}